\documentclass[aps,preprintnumbers,showpacs,nofootinbib,a4paper]{revtex4}
\usepackage{subfig}
\usepackage{amsmath}
\usepackage{amsfonts}
\usepackage{amssymb}
\usepackage{mathptmx}
\usepackage{slashed}
\usepackage{latexsym}
\usepackage{graphicx}
\usepackage{color}
\usepackage{float}
\usepackage{multirow}
\newcommand{\newc}{\newcommand}
\newc{\beg}{\begin}
\newc{\beq}{\beg{equation}}
\newc{\eeq}{\end{equation}}
\newc{\bea}{\beg{eqnarray}}
\newc{\eea}{\end{eqnarray}}
\newc{\m}{\mathcal}
\newc{\nn}{\nonumber}
\newc{\diag}[1]{\ensuremath{{\rm diag} \left( #1 \right)}}
\newc{\ea}{\end{eqnarray}}
\newc{\barr}{\begin{eqnarray}}
\newc{\earr}{\end{eqnarray}}
\newc{\pa}{\partial}
\newc{\bfe}{{\bf \ell}}
\newc{\alp}{\alpha}
\newc{\gam}{\gamma}
\newc{\bfeta}{{\bf \eta}}
\newc{\bfxi}{{\bf \xi}}
\newc{\del}{\delta}
\newc{\eps}{\epsilon}
\newc{\lam}{\lambda}
\newc{\sig}{\sigma}
\newc{\ups}{\upsilon}
\newc{\ome}{\omega}
\newc{\ra}{\rightarrow}
\newc{\ov}{\overrightarrow}
\newc{\la}{\langle}
\newc{\rg}{\rangle}
\newc{\im}{\imath}
\begin{document}

\title{ Neutron disappearance inside the nucleus}

\author{ H. Ejiri$^{1}$, J.D. Vergados$^{2}$}

\affiliation{$^1$RCNP, Osaka University, Osaka, 567-0047, Japan} 
\affiliation{$^2$ University of Ioannina, Ioannina, Gr 451 10, Greece \footnote{e-mail address: vegados@uoi.gr} }
\vspace{0.5cm}
\begin{abstract}
 We consider the possibility that a neutron may disappear inside the nucleus, which will demonstrate the existence of baryon violating $\Delta B=1$ interactions. It has recently  been proposed that such a process may have an effect on the free neutron decay life time. We evaluate the widths for $n\rightarrow \chi$ and $n\rightarrow \chi \gamma$, with $\chi$ being a light dark matter particle emitted by a loosely bound neutron in various light nuclei. We find that, assuming a mass $m_{\chi}$ close to 938 MeV, the obtained width for  $n\rightarrow \chi$ in $^{11}$Be is much larger than the observed beta decay width. This suggests a severe limit on the possible decay channel of $n \rightarrow \chi \gamma $ for free neutron.

\end{abstract}
\date{\today}
\maketitle



\section{Introduction}
The neutron is one of the 
building blocks of matter.  Without it, complex
atomic nuclei simply would not have formed. Although
the neutron was discovered over eighty years ago  and has
been studied intensively thereafter, its precise lifetime is still
an open question \cite{WietGreene11,GreeneGelt16}. 
There are two qualitatively different types of direct neutron
lifetime measurements: bottle and beam experiments.
In the first method one obtains \cite{Patrignani15}:
\beq
t_n(\mbox{bottle})=(879.6\pm 0.6)\mbox{s}.
\eeq
In the second, beam method, the result as given by 
PDG average \cite{ByrDaw96,YDGGLNSW13} is 
\beq
\tau_n(\mbox{beam})=(888\pm 2.0)\mbox{s}.
\eeq 
The discrepancy between the two results is $4.0\sigma$. 

This suggests
that either one of the measurement methods suffers from
an uncontrolled systematic error, or there is a physics  reason of
why the two methods give different results, involving very interesting physics.
It is interesting to note that in the beam experiment the life time is longer.

The above facts were known to the authors of a recent paper \cite{FornalBej} and, in particular,  the problem of  the  discrepancy   between the two experimental measurements of the neutron decay lifetime has been addressed in their work.
They noted that, since in the "`beam"'  experiment  the result is obtained by studying   the decay, the lifetime
they measure is related to the actual neutron lifetime by
\beq
\tau_n(\mbox{beam}) =\frac{\tau_n}{\mbox{Br}(n\rightarrow p+\mbox{anything})}
\eeq
These authors suggest that the discrepancy can be explained by considering an extra channel in the beam experiment, which involves the emission of a dark fermion particle $\chi$, which goes undetected. Then they proposed a model which can give a branching ratio of $1\%$ to this new channel, while the standard channel covers only $99\%$, thus settling the issue.  It is, however, necessary in their treatment  to assume that this new particle is a neutral Dirac like fermion with a mass slightly lighter than that of the neutron. If it were a Majorana like particle this mechanism could lead to neutron-antineutron oscillations, thus  being in conflict with the non observation of such oscillations.

 For a free neutron this cannot occur without the emission of another particle, e.g. a photon to conserve energy momentum, which at the same time  provides an interesting experimental signature for the suggested mechanism.

 Since the emitted particle is assumed not to carry any baryon number \cite{FornalBej}, this scenario is very interesting, since if true, it will demonstrate the existence of baryon number violating $\Delta B=1$ interactions. This scenario seems, however, to be excluded from astrophysical data involving neutron stars \cite{Motta18}, \cite{McKeen18}, \cite{Baym18}. Such arguments, however, do not apply, if there
exists a Coulomb repulsion between the dark fermions mediated by a dark photon \cite{ClineCorn18}. In any case the neutron star arguments 
apply only to neutrons  bound by gravity and not by strong interactions.

The neutron in the nucleus seems to  behave differently, due to the nuclear binding. In certain cases it decays like in the $\beta$ decay, but the produced proton cannot escape due to nuclear binding, while a daughter nucleus appears with its charge increased by one unit. Decays of well-bound nucleons (neutrons) into invisible particles have been searched 
by measuring $\gamma$ rays long time ago \cite{Ejiriphoa93,Ejiriphob94}. In the model considered above the produced dark matter particle $\chi$,  interacting very weakly, can escape. In this case energy-momentum can be conserved without the emission of additional particles, like the photon, and the decay width is expected to be much larger.

 In this work we will consider the possibility that such a process may occur inside the nucleus:
\beq
A(N,Z)\rightarrow A(N-1,Z)^* +\chi
\label{Eq:NuclChi}
\eeq
where $\chi$ is the dark matter fermion. Since $\chi$ is also supposed to be produced in the decay of the free neutron, it must  be lighter than the neutron. In fact to explain the neutron life discrepancy it was necessary to assume \cite{FornalBej} its mass must be less than but very close to that of the neutron $m_n=939.565$:
\beq
937.900\mbox{ MeV}<m_{\chi}<938.783\mbox{ MeV},
\label{Eq:masschi}
\eeq
 where $m_{\chi} $ is the dark particle mass. The lower and upper bounds come from the stability of $^9$Be and  the fact that the  decay  $\chi\rightarrow p + e^- + \bar{\nu_e}$ is forbidden, respectively. Note that the lower bound corresponds to $m_n-B_n$ with $B_n=1.67$ MeV being the neutron binding energy of $^9$Be.

While our paper was in preparation a related paper appeared \cite{CzarMarSir18}, but it was based on experimental information. In any case  we can not see how the factor $g_A$ they employ could appear in a particle model like ours, with the baryon violating interaction  mediated by scalar particles.

Before proceeding to a further  study of the baryon number violating process, we will consider some facts about the nuclei, which can serve as targets: Nuclei that need to be studied experimentally should have a  loosely bound neutron with $B_n\approx $0.5 MeV - 1 MeV  to decay to $\chi$, as in Eqs. (\ref{Eq:NuclChi}) and (\ref{Eq:masschi}), and a long $\beta $ decay half-life of the order of seconds  in order to make the possible decay with much shorter half-life well separated from the $\beta $ decay. The long half life is necessary to make the decay to $\chi$ visible. This way we have two  possible candidates, $^{11}$Be and $^{15}$C. Thus, we will mainly discuss the  possible decay of a bound neutron to the ground state in the residual nucleus in these two cases.

 $^{11}$Be has already been  studied experimentally, see, e.g., ref. \cite{BeHalo12} and references there in. Its ground state is $1/2^+$ and the first excited state is $1/2^-$.  It seems that the  $1/2^+$ is an  unusual ordering from the point of view of the simple shell model. It has also been studied both  in the context of effective field theory \cite{HamPhil11}, focusing on its electric properties, and the no-core
shell model with continuum (NCSMC) type calculations \cite{CNRDEQ16}. The later admittedly not, strictly speaking, an ab initio calculation, has succeeded in getting  the inversion right. Thus the  $1/2^+$ is composed of a major component  of the form  $^{10}\mbox{Be g.s}\otimes 1s_{1/2}(n)$ and, possibly,  of another one of the type  $^{10}\mbox{Be}2^+\otimes 0d_{5/2}(n)$. The first can decay into the ground state (g.s.) of $^{10}$Be.
 
On the theoretical side there have been a lot of additional studies  \cite{BayeCapGol07,CapNun06,CapNun07}. Variational Model approach \cite{OtsFukSag93} as well as models which vary the single 
energies via vibrational  and rotational
core couplings, succeed in reproducing the needed level inversion in a rather systematic
manner. One can say that a common feature for the success of these models is
the inclusion of core excitation. On the other hand ab initio No-Core Shell
Model calculations \cite{Caurier05}  have not  been able to reproduce this
level inversion. Anyway it is considered a success that a significant drop in the energy of
the 1/2+ state in $^{11}$Be is reported with enlarging the model
space. 

Relevant useful experimental as well as theoretical information can be found in a previous work \cite{Winfield01}, in particular the fact that the relevant spectroscopic factor for the first component  is about (80$\pm 10) \%$.

\section{The formalism}
\subsection{Neutron bound wave functions}
 We will consider the neutron as a bound state  of three  quarks in a color singlet s-state . The orbital part  of the form:
\beq
\Psi(R,\xi,\eta)=\Phi(R)\psi_{0s}(\bfxi)\psi_{0s}(\bfeta). 
\eeq
The $\psi_{0s}(\bfxi)$ and $\psi_{0s}(\bfeta)$ are bound wave functions dependent on the relative internal variables, which, for simplicity, we will assume  to be of the $0s$ harmonic oscillator type, so that one can easily separate  out the internal coordinates.
Thus
\beq
\psi_{0s}(\xi)=\sqrt{\frac{1}{\pi \sqrt{\pi}}}\frac{1}{(b_N)^{3/2}}e^{-\frac{\xi^2}{2 b_N^2}},\,\, \xi=\frac{1}{\sqrt{2}}({\bf x}_1-{\bf x}_2).
\eeq
Similarly
\beq
\psi_{0s}(\eta)=\sqrt{\frac{1}{\pi\sqrt{\pi}}}\frac{1}{(b_N)^{3/2}}e^{-\frac{\eta^2}{2 b_N^2}},\,\, \eta=\frac{1}{\sqrt{6}}({\bf x}_1+{\bf x}_2-2{\bf x}_3).
\eeq
The center of mass coordinate is taken to be:
\beq
{\bf R}=\frac{1}{\sqrt{3}}({\bf x}_1+{\bf x}_2+{\bf x}_3)
\eeq
with ${\bf x }_i,\,i=1,2,3$ the quark coordinates and $b_N$  the nucleon size parameter related to the nucleon radius $R_N$ via the relation  $R^2_N=(3/2)b^2_N$. 
The functions $\phi(\bfxi)$ and $\phi(\bfeta)$ are  normalized in the usual way.
\subsection{The amplitude for neutron decay to a dark mater fermion}
\begin{figure}[h]
\begin{center}
\subfloat
{
\includegraphics[width=0.3\textwidth]{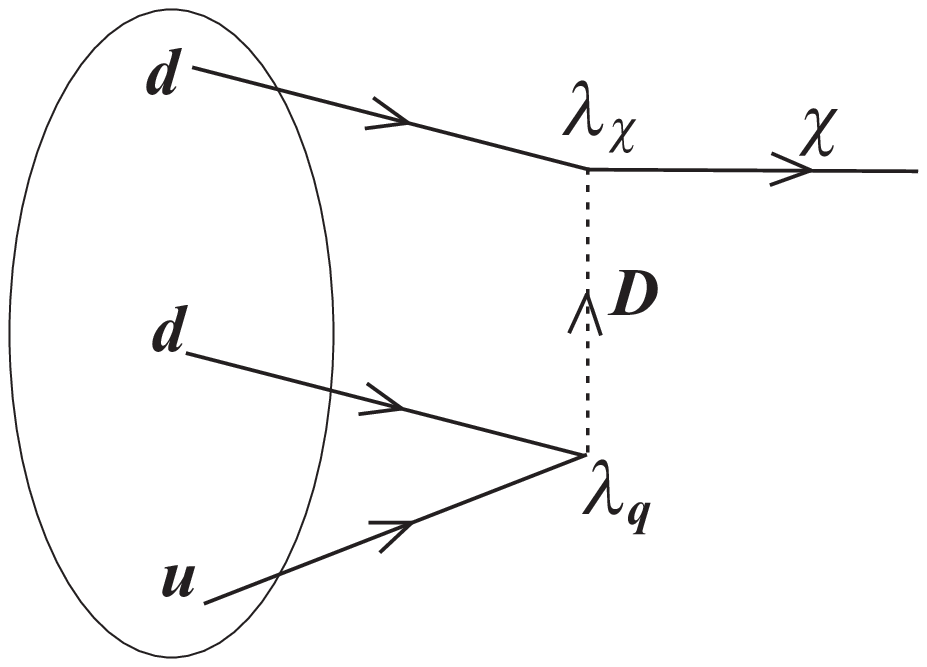}
}
{
\includegraphics[width=0.4\textwidth]{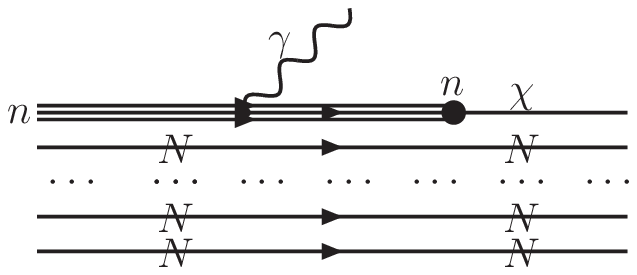}
}
\\
\caption{  Left panel: A dark matter colorless Dirac fermion is emitted  from a neutron (n) viewed as a bound state of three quarks. At the quark level the baryon number violating process is mediated by a colored scalar field D. Right panel: A bound neutron in a nucleus is converted into a dark matter particle. Before this happens it may also emit a photon, which in the nucleus is not necessary. The other  $A-1$ nucleons ($N$) do not participate in the process}
\label{Fig:ndecay}
\end{center}
\end{figure}

The process derived from  the model of \cite{FornalBej} is exhibited in Fig. \ref{Fig:ndecay}
The orbital part of the amplitude associated with this process takes the form:
\beq
{\cal M}=\frac{\lambda_q\lambda_{\chi}}{m_{\Phi}^2} \psi_{0s}(\bfxi)\psi_{0s}(\bfeta) (2 \pi )^3 \delta({\bf p}_1+{\bf p}_2+{\bf p}_3-{\bf q}),
\eeq
where $\lambda_q$ and $\lambda_{\chi}$ are the Yukawa couplings of the scalar field $D$ as shown in Fig. \ref{Fig:ndecay}, with $m_{\Phi}$ the mass of the scalar.  ${\bf p}_i,\,i=1,\,2,\,3$ are the quark momenta and ${\bf q}$ the momentum of the outgoing momentum of the dark particle.

The Fourier transform of the amplitude in coordinate space becomes: 
\beq
{\cal M}=\frac{1}{(2 \pi)^6}\frac{\lambda_q\lambda_{\chi}}{m_{\Phi}^2} \psi_{0s}(\bfxi)\psi_{0s}(\bfeta)\int d^3{\bf p}_1 \int d^3{\bf p}_2 \int d^3{\bf p}_3 e^{i {\bf p}_1.{\bf x}_1}e^{i {\bf p}_2.{\bf x}_2}e^{i {\bf p}_3.{\bf x}_3}\delta({\bf p}_1+{\bf p}_2+{\bf p}_3-{\bf q})
\eeq
or
\beq
{\cal M}=\frac{\lambda_q\lambda_{\chi}}{m_{\Phi}^2} \psi_{0s}(\bfxi)\psi_{0s}(\bfeta) \delta({\bf x}_1-{\bf x}_3)\delta({\bf x}_2-{\bf x}_3)e^{i {\bf q}.{\bf x}_3}
\eeq
or
\beq
{\cal M}=\frac{\lambda_q\lambda_{\chi}}{m_{\Phi}^2} \psi_{0s}(\bfxi)\psi_{0s}(\bfeta) \delta\left (\sqrt{2}\bfxi\right )\delta\left (\sqrt{\frac{3}{2}}\bfeta\right )e^{i \frac{\bf q}{\sqrt{3}}.(-\sqrt{2} \bfeta+\bf R)}.
\eeq
Thus we get:
\beq
{\cal M}=\frac{1}{2 \sqrt{2}}\frac{2\sqrt{2}}{3 \sqrt{3}}\frac{\lambda_q\lambda_{\chi}}{m_{\Phi}^2} |\psi_(0)|^2 e^{i \frac{\bf q}{\sqrt{3}}.{\bf R}}.
\eeq
\subsection{The spin-isospin-color factor}
We must now consider the internal degrees, i.e.  spin-isospin-color factor. We will employ group theoretical techniques.  Indeed the neutron isospin $I=1/2$ implies that the corresponding $SU(2)$ symmetry corresponds to a Young tableaux $[f^{I}]=[2,1]$. Since the orbital part for $0s$ quarks is completely symmetric, the spin color part for a completely antisymmetric neutron wave function must correspond to the associated  Young Tableaux, i.e. that obtained from the previous by interchanging rows and columns, which in this case is  $[f^{sc}]=[2,1]$ again, which, of course,  contains \cite{SOSTRO} a color singlet with spin 1/2, $[2,1](0,0)1/2$. The two quarks appearing in the process of Fig. \ref{Fig:ndecay}, symbolically indicated a $ud$,  must be in a spin zero state, in order to couple to a scalar particle, and  must be in a color anti-triplet. Thus they must be in an isospin zero state, which means that the isospin CFP (coefficient of fractional parentage) is unity. The spin-color one particle CFP connecting one quark to the two quark color anti-triplet, i.e. $[2](0,1)s=0$,  is also unity \cite{SOSTRO}. So the corresponding matrix element is
$ \sqrt{3} \sqrt{\frac{\mbox{dim}[2]}{\mbox{dim}[2,1]}}=\sqrt{\frac{3}{2}}$, where the first number corresponds to the number of quarks in the neutron and the second comes from the dimensions of the relevant representations of the symmetric group.
Thus finally
\beq
{\cal M}=\kappa_{\mbox{{\tiny scale}}}e^{i \frac{\bf q}{\sqrt{3}}.{\bf R}},\,\,\kappa_{\mbox{{\tiny scale}}}=\frac{1}{3\sqrt{2}}\epsilon, \epsilon=|\psi(0)|^2 \frac{\lambda_q\lambda_{\chi}}{m^2_{\Phi}},
\eeq
where 
\beq
|\psi(0)|^2 = \frac{1}{\pi \sqrt{\pi}}\frac{1}{b^3_N},\,\,b_N=\sqrt{\frac{2}{3}}R_N
\eeq
is the baryon density at the origin.
\subsection{Fixing the nucleon size parameter} 
 For a typical value of $R_N=0.8$ fm we obtain a value of $ |\psi(0)|^2=0.005$ GeV$^3$. This is a bit smaller than the analogous parameter $\beta$ employed in the free neutron decay \cite{FornalBej}, i.e.  $\beta= 0.014$ GeV$^3$, obtained recently from lattice computation of proton decay matrix elements \cite{LatticeQCD17}. We will adopt the value of $ b_N=0.5$ fm  to be consistent with the larger value of  $\beta= 0.014$ GeV$^3$.
Thus
\beq
\kappa_{\mbox{{\tiny scale}}}=\frac{1}{3 \sqrt{2}}\frac{1}{\pi \sqrt{\pi}}\frac{1}{b_N}\frac{\lambda_q\lambda_{\chi}}{(b_N m_{\Phi})^2}.
\eeq
We find it convenient to indicate  the size of the baryon number  violating neutron conversion  to a dark mater fermion  by the dimensionless quantity  
$s=\left (\frac{\lambda_q\lambda_{\chi}}{(b_N m_{\Phi})^2} \right )^2$ instead of the dimensionful parameter  $\epsilon^2$ used in   \cite{FornalBej}. Using the value of $(\lambda_q \lambda_{\chi})/m^2_{\Phi}=6.7\times 10^{-6}$TeV$^{-2}$, i.e. the one   employed in the free nucleon exotic decay to dark matter \cite{FornalBej}, we find that both are indeed very small:
$$ s\approx 1.5\times 10^{-24}\mbox{ or } \epsilon^2=8.8 \times 10^{-27}\mbox{GeV}^2$$
Either one can be used  in bound as well as in free nucleon decay.
\section{Expression for the decay width}
\begin{figure}[h]
\begin{center}
\includegraphics[width=0.8\textwidth,height=0.5\textwidth]{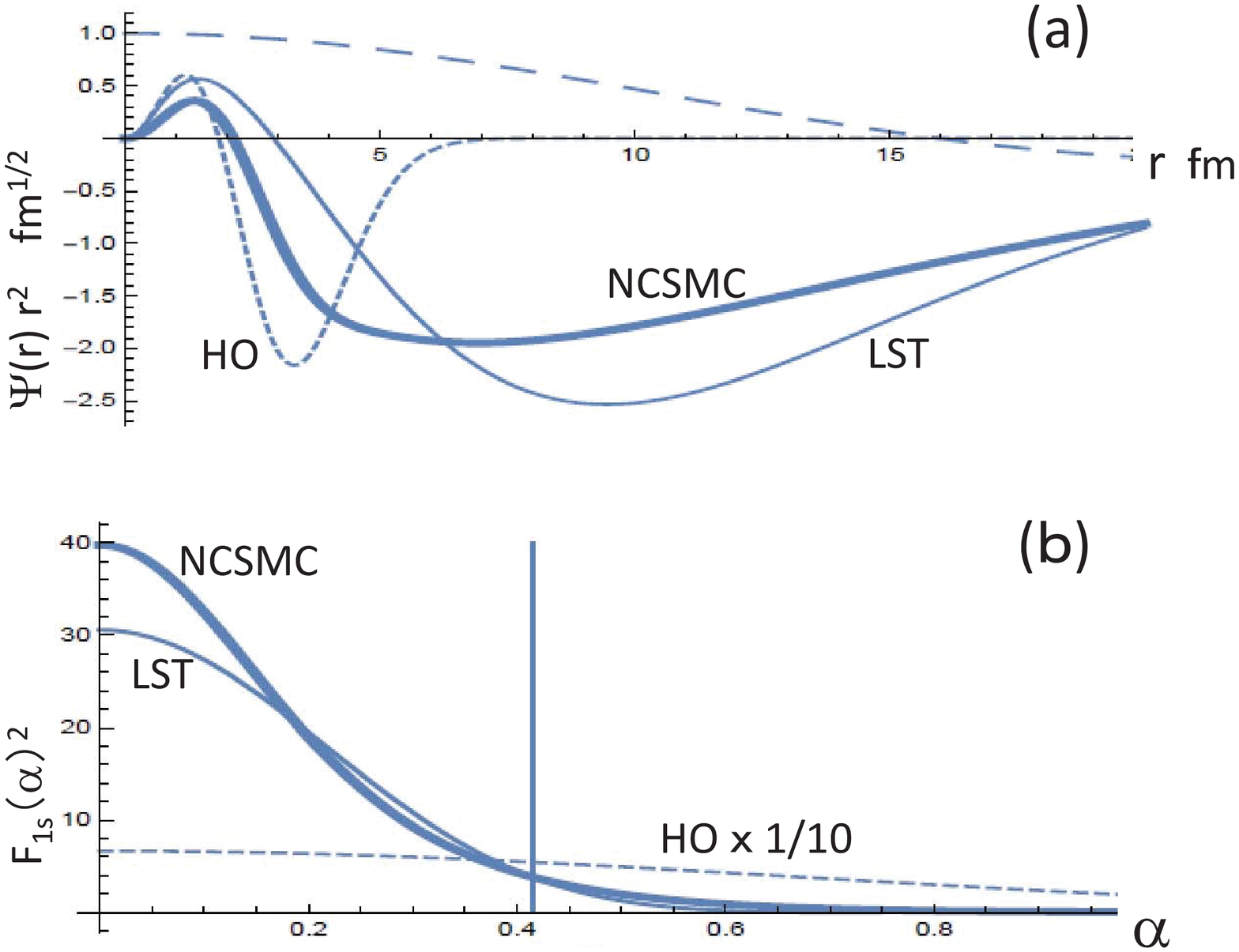}
\end{center}
\caption{ In the top panel (a) we plot  the expression $r^2  \Psi(r)$, with $ \Psi(r)$ the wave function of the loosely bound neutron in $^{11}$Be as follows: solid line corresponds to the empirical local scale transformations (LST), thick solid line to the  NCSMC  and the dashed line to the harmonic oscillator case. One clearly sees the importance of the sign change in the nuclear  wave function.  We also exhibit the behavior of the spherical Bessel function  $j_0(q r)$, which is, of course, dimensionless, 
  for  a small momentum transfer of interest to us, i.e.  $q=0.2$ fm$^{-1}$ $\Leftrightarrow m_n-B-m_{\chi}\approx$  0.8 MeV, (long dashed curve),  merely  to indicate that it does not make real   oscillations. It eventually changes sign, but all nuclear wave functions are small there. In the bottom panel (b) we exhibit the square of the form factors associated with the $1s$ state of $^{11}$Be for the  HO oscillator case (dashed line),  the LST wave function  (solid line) and the NCSMC (thick solid line) are exhibited. Note that the presented here  HO results have been multiplied by a factor of 10, to make the dashed line visible. The vertical line corresponds to  value of $\alpha$ associated with  the energy $m_n-B-m_{\chi}\approx$  0.8 MeV, which was used in the present calculation.}
\label{Fig:facndecay}
\end{figure}

The decay width is given by the expression:
\beq
d \Gamma=\frac{1}{(2 \pi)^2}d^3 {\bf q}d^3 {\bf P}_A \delta({\bf q}+{\bf P}_A)\delta(\Delta-E_x +m_n-m_{\chi}-T)|\langle \mbox{ME}\rangle|^2
\eeq
where  ${\bf P}_A$ and ${\bf q}$  are the momenta of the final nucleus and the outgoing dark matter  particle $\chi$ respectively, with the latter's mass being $m_{\chi}$ and its kinetic energy $T$. $\Delta$ is the difference of the ground state energies of the nuclei involved with the neutron mass separated out and $E_x$ the excitation energy of the populated final nuclear state. Finally  
ME (matrix element) is the invariant amplitude which will be  been given in in Eq. (\ref{Eq:InvAmpSq}) of the appendix.
Thus
\beq
\Gamma=\frac{1}{\pi}\sqrt{2 m_{\chi}(\Delta-E_x +m_n-m_{\chi})}m_{\chi}|\langle \mbox{ME}\rangle|^2\approx
\frac{1}{\pi}\sqrt{2 m_{n}(\Delta-E_x+m_n-m_{\chi})}m_{n} |\langle \mbox{ME}\rangle|^2
\eeq
or, indicating by $j(n)$ the   angular momentum of the loosely bound neutron, we obtain:
\beq
\Gamma=\frac{16}{9\pi^3}\frac{a_N^3}{b_N^2}\left ( \frac{\lambda_q\lambda_{\chi}}{(b_N m_{\Phi})^2}\right )^2\sqrt{2 m_{n}(\Delta-E_x+m_n-m_{\chi})}m_{n}\langle A(N-1,Z)J_f;j(n);A(N,Z)\rangle^2 2 f^2_{j,\ell}\left (F_{n_r\ell} (\alpha) \right )^2,
\label{Eq:totalwidth}
\eeq
with $F_{n_r\ell} (\alpha)$ the relevant form factor defined by Eq. (\ref{Eq:NucFormFac}) and $\alpha$ given by Eq. (\ref{Eq:AlphaTran}) below.\\
The function  $f^2_{j,\ell}$ is an angular momentum dimension coefficient, which in our case is trivial, i.e. $f^2_{j,\ell}=1/2$, see Eq. (\ref{Eq:fjl}).
The form factors for harmonic oscillator wave functions have been obtained analytically in the appendix. We will, however, illustrate  their behavior in Fig. \ref{Fig:facndecay}. 
The needed nuclear CFP's can be calculated in a shell model treatment or perhaps they can be extracted  from other experiments as mentioned earlier.

 We note that in the interesting case for $1s$ orbital, the form factor   is sensitive to the nuclear model employed,
since this  form factor essentially involves the overlap of a bound nucleon wave function with a plane wave. The result  is particularly sensitive to the parameters for nucleon wave functions that have nodes, like the $1s$ orbit, since at some point there appears a change in sign. Thus, due to the $r^2$ factor appearing in the integral, the behavior at large $r$ becomes crucial.  One might have thought that large additional suppression may be due to the oscillations of the spherical Bessel function $j_0(q r)$, see Eq. (\ref{Eq:Jme}). In the kinematic region of interest to us, this is not the case, since the  spherical Bessel function does not make many oscillations, see Fig. \ref{Fig:facndecay}.  It merely changes sign at around 16 fm, but all nuclear wave functions are small there.\\
 In view of the above, the harmonic oscillator description, however, may not be satisfactory in describing the conversion of a nucleon in the nucleus into a dark matter particle, since a neutron with a binding energy of about 1 MeV may extend further than that prescribed by  the harmonic oscillator (HO)  model. 

One therefore has to modify these wave functions. To this end we recall that  wave functions obtained in the context of  no core shell model with continuum (NCSMC)  \cite{CNRDEQ16} as well as in the approach of local scale transformations
 (LST) \cite{karaAmos05}, have been found useful in the study  of extended matter distributions in nuclei. So we are going to employ both of them in the present calculation. In the first approach  using the NCSMC wave functions  provided to us by Navratil \cite{CNRDEQ16}. In the second approach, using the techniques the of local scale transformation
(LST) of Karataglidis and  Amos \cite{karaAmos05}, we obtained the needed wave functions as discussed in section 
\ref{sec:Localcale}. The main relevant  results are presented in Fig. \ref{Fig:facndecay}.


It should mentioned that the  the harmonic oscillator size parameter employed has been obtained in the usual way:
\beq
a_N=\frac{\hbar c}{\sqrt{\hbar \omega m_nc^2}},\,\hbar \omega=41 A^{-1/3} \mbox{MeV}
\eeq
This leads to the value of $a_N=$1.5 fm for the A=11 nucleus and  $1.55$ for the A=15 system.\\ The LST model is not sensitive to this parameter, since the long range behavior of the wave function is of exponential form with a range related to a neutron  energy  \cite{karaAmos05}  as discussed in the appendix. For short distances we  used the above $a_N$ with the transition occurring at the same  distance $a_N$.\\
In the NCSMC model a somewhat smaller size parameter, 1.44 fm, was employed by Navratil \cite{CNRDEQ16}, but the long range behavior of his wave function is not sensitive to this choice.

We find that, for a binding  energy of 0.51 MeV and the choice of $m=4$, see  section \ref{sec:Localcale}, the LST calculations for $1s$  wave functions yield an  enhancement of the form factor $\left (F_{n_r\ell} (\alpha) \right )^2$ by about 7.0 times, compared to those obtained with   harmonic oscillator wave functions, for a value of $\alpha$ around 0.4 (for the definition of $\alpha $ see Eq. (\ref{Eq:AlphaTran}) below). For $m=2$ the results change slightly, this increases to 7.6. So we will adopt here the value of $m=4$.    In the case of the  NCSMC for $^{11}$Be we get an enhancement of the form actor of about 7.1 compared to that obtained with  the HO. Thus  for $^{11}$Be the LST and  NCSMC are in agreement. The reason that the enhancement is not very large is the fact the parameter $\alpha$, proportional to the momentum transfer, is quite large (see table \ref{Eq:alphafalpha}), for $\delta E=m_n-B- m_{\chi}$=0.8. The  HO form factor does not fall with $\alpha$ as fast as the other two do.

 We now come to the second candidate, namely $^{15}$C. Here the spectroscopic factor is determined experimentally to be $(0.75\pm 0.15)\%$ and we will adopt the value 0.75 \cite{Gross75}  for the 1s neutron. We will use  $a_N=1.55$ both for the HO and the LST calculations. The latter is not sensitive to  this value, since, as we have mentioned, the long range behavior of the wave function is dominated  by the the neutron binding energy, which is found in standard tables \cite{Isotopes} $B=1.218$ MeV. Because of this larger binding we get $\delta E=m_n-B- m_{\chi}$=0.092. The   width is enhanced by a factor of  21 compared to that of the HO. The reason that this enhancement is now large is the fact the value of $\alpha$ is small (see table \ref{Eq:alphafalpha}).
 We are not sure of  whether a similar enhancement exists in the case of the NCSMC, so we prefer to leave  blank the appropriate place in table \ref{tab:outdata}.

\section{Some results}
  
 To simplify matters let  us for the moment  assume the same nuclear size parameter, e.g.    $a_N=$1.5 fm, for the nuclei considered here  and a nucleon size parameter of $b_N=0.5$ fm. Let us also take the value of $(\lambda_q \lambda_{\chi})/m^2_{\Phi}=6.7\times 10^{-6}$ TeV$^{-2}$, i.e. the one   employed in the  sister free nucleon exotic decay to dark matter \cite{FornalBej}. 
In the special case of halo nuclei  considered here $\Delta=-B$ with $B>0$ the binding energy of the decaying neutron and transitions to the ground state, $E_x=0$,
 Eq. (\ref{Eq:totalwidth}) becomes:
\beq
\Gamma= 1.0 \times 10^{-16}\mbox{eV} \sqrt{\frac{m_n-B-m_{\chi}}{\mbox{1 MeV}}} g(\alpha),\,\,g(\alpha)=\langle A(N-1,Z)J_f;j(n);A(N,Z)\rangle^2 f^2_{j,\ell}\left (F_{n_r\ell} (\alpha) \right )^2
\label{Eq:widthbound}
\eeq
with $j(n)$ indicating the state of the loosely bound neutron and 
 \beq
\alpha=\sqrt{2}\frac{\sqrt{2 m_n(m_n-B -m_{\chi})} a_N}{\hbar c}.
\label{Eq:AlphaTran}
\eeq
In other words the dimensionless parameter $\alpha$ is proportional to the momentum $ \sqrt{2 m_n(m_n-B -m_{\chi})}$ of the emitted dark matter particle $\chi$. 

Let us consider the invisible decay of a loosely-bound neutron in a nucleus. It turns out there exist several nuclei  to be considered as given in table \ref{tab:outdata}.
 The deuteron and $^{9}$Be are stable isotopes with the half-life much longer than  that of the universe. 
Then their small limits on the widths exclude the n decays to $\chi$ with $m_{\chi} \le $ 937.3 MeV and  $m_{\chi} \le$ 937.9 MeV, respectively \cite{FornalBej}. 
  Then the $\chi$ mass  region is limited in a narrow region of 937.9 MeV 
$\le m_{\chi}\le$ 939.56 MeV. 

We are now in position to estimate the widths in the case of the nuclei of interest. The relevant nuclear parameters are contained in the   quantity $g(\alpha)$  and are presented in table \ref{tab:outdata}.  Regarding the nuclear input in the case of $^{11}\mbox{Be}$ we used an average spectroscopic factor  $\langle A(N-1,)J_f;j(n);A(N,Z)\rangle^2= 0.85$ \cite{Winfield01} for a transition to the ground state. Also in the case of the 1s HO, NCSMC and LST wave function for various values of $ m_n-m_\chi -B$  we find the  form factors given in table \ref{Eq:alphafalpha}.
\begin{table}[htbp]
\begin{center}
\caption{ The form factors squared considered in this work.The first  two values of $\alpha$ correspond to $^{11}$Be and the last two  correspond to $^{15}$C.
}
\label{Eq:alphafalpha}
$$
\begin{array}{|c|c|c|c|c|}
\hline
&^{11}\mbox{Be}&^{11}\mbox{Be}&^{15}\mbox{C}&^{15}\mbox{C}\\
\hline
m_n-m_\chi -B&0.8\mbox{ MeV}&1.0\mbox{ MeV}&0.092\mbox{ MeV}&0.115\mbox{ MeV}\\
\hline \alpha&0.415&0.464&0.140&0.158\\
\hline 
F^2_{1s}(\alpha)\mbox{ (HO)}&0.541&0.514&0.649&0.646\\
F^2_{1s}(\alpha)\mbox{ (NCSMC)}&3.82&2.60&-&-\\
F^2_{1s}(\alpha)\mbox{ (LST)}&3.79&2.08&14.1&13.5\\
\hline \end{array}
$$
\end{center}
\end{table}

\begin{figure}[hbtp]
\begin{center}
\includegraphics [width=0.7 \textwidth]{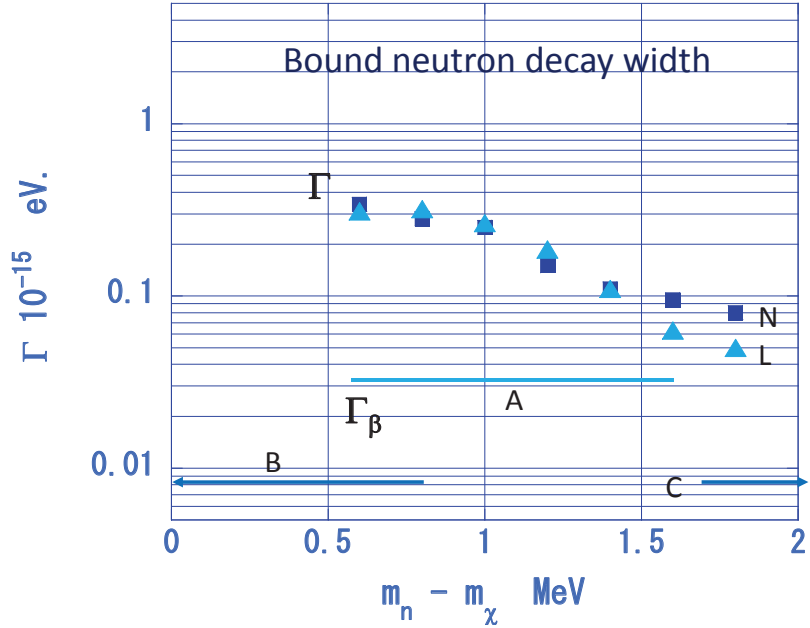}
\caption{The experimental decay width $\Gamma _{\beta}$ for $^{11}$Be (thick line), the evaluated via  Eq. (\ref{Eq:widthbound}) width $\Gamma _{\chi}$ (squares for model NCSMC, marked as N, and triangles  for LST, marked as L) , and
 the excluded mass-regions  with
A: by the $^{11}$Be  decay width, B: by the decay to proton, and C: by the $^{9}$Be,   
respectively. The size of the evaluated value reflects the effect of the uncertainty of 15$\%$ in the spectroscopic factor.
The uncertainty of around 1 $\%$ in the experimental error in the half-life is within the thickness of the line.
\label{fig:figbe11}}
\end{center}
\end{figure}

The decay half-life of $^{11}$Be is measured to be 13.81 $\pm 0.08$ s by counting the decay particle as a function of the time \cite{AlbEng70}. The width is 3.3 $\times $ 10$^{-17}$
 eV, which is much smaller than the evaluated width given by (\ref{Eq:widthbound}),
 as shown in Fig. \ref{fig:figbe11}.
Note that the observed $^{11}$Be width is smaller by a factor of about $10-3$ compared  to the  evaluated widths  based of the realistic LST and NCSMC form factors  
in the region of $m_n-m_{\chi}=0.8 - 1.6$ MeV.

Thus the decay to 
the DM with $m_{\chi} \le $ 939.0 MeV is excluded. Note that this limit is very insensitive to the nuclear structure coefficient $g$ given in Eq.  (\ref{Eq:widthbound}). The lighter $\chi$ with $m_{\chi} \le$ 937.9 MeV is excluded by the  long-lived $^{9}$Be \cite{FornalBej}.   
 On the other hand the heavier DM with $m_{\chi} \ge$ 938.78 MeV is excluded since the decay of $\chi \rightarrow p+e^- +\bar{\nu}_e$ is forbidden \cite{FornalBej}.

Thus the decay of the bound neutron is excluded, and thus the suggested decay of the free neutron to $\chi+\gamma$ is not likely.  
 Note that $\chi$ is assumed not to decay to $p+e^-+\bar{\nu}_e$ \cite{FornalBej} and thus, if we allow unstable $\chi$, the mass region of 939.565-939.0 is not excluded.

The  measured width of 1.9 10$^{-16}$ eV for $^{15}$C  is of the same order of magnitude as the evaluated ones and, thus, disfavors the possibility of DM with $m_{\chi} \le$ 938.3 MeV. 
 Recently  n decays in  various nuclei have been discussed \cite{pfu18}.

\begin{table}[htbp]
\begin{center}
 \caption{ The expected widths for neutron decay into a dark matter particle $\chi$  inside a   nucleus  for the indicated values of $\delta E=m_n-B- m_{\chi}$. S stands for stable isotope. We present results obtained  with the  HO  shell model wave function as well as  for the NCSMC and LST models discussed in the text. The last two models may be more appropriate for halo type nuclei.  }
\label{tab:outdata}
$$
\begin{array}{|c|c|c|c|c|c|c|c|c|c|c|c|}
\hline
\mbox{nucleus}&B_n&J^{\pi}&j_n&T_{1/2}\mbox{[s]}&\delta E &g(\alpha)&\Gamma[10^{-16}\mbox{eV]}&g(\alpha)&\Gamma[10^{-16}\mbox{eV]}&g(\alpha)&\Gamma[10^{-16}\mbox{eV]}\\&\mbox{[MeV]}&&&&\mbox{[MeV]}&\mbox{(HO)}&\mbox{(HO)}&\mbox{(NCSMC)}&\mbox{(NCSMC)}&\mbox{(LST)}&\mbox{(LST)}\\
\hline
\hline
^2\mbox{H}&2.225&1^+&0s_{1/2}&S&-&-&-&-&-&-&-\\
\hline
^9\mbox{Be}&1.665&(3/2)^-&0p_{3/2}&S&-&-&-&-&-&-&-\\
\hline
^{11}\mbox{Be}&0.504&(1/2)^+&1s_{1/2}&13.8&0.80&0.221&0.20&1.63&1.4&1.61&1.4\\
\hline
^{15}\mbox{C}&1.218&(1/2)^+&1s_{1/2}&2.45&0.092&0.243&0.073&-&-&5.1&1.6\\
\hline
\end{array}
$$
\end{center}
\end{table}

As we have mentioned the radiation of a photon by the bound decaying neutron is not needed. We have , however, estimated in the appendix the branching ratio for such a process.
For the most interesting case of $1s_{1/2}$ neutron we obtain:
\beq
\frac{\Gamma \left (J_i\rightarrow J_f \chi \gamma\right)}{\Gamma \left (J_i\rightarrow J_f \chi \right)}\approx 2.4\times 10^{-7}\frac{(m_n-B-m_{\chi})^2}{1 \mbox{MeV}^2}.
\eeq

The width for the radiative free neutron decay is given \cite{FornalBej} by:
\beq
\Gamma(n\rightarrow  \gamma \chi)=\frac{g^2_s}{8 \pi} 4 \pi \alpha \left( 1-\frac{m^2_{\chi}}{m^2_n}\right )^3 \frac{\epsilon^2}{(m_n-m_{\chi})^2} m_n
\eeq
where $\epsilon =9.38\times 10^{-14}$ GeV and $g_s$ is the neutron g-factor, $g_s=-3.80$. Here $\alpha$ is the fine structure constant, not to be confused with that of Eq. (\ref{Eq:AlphaTran}).  Then one obtains:
\beq
\Gamma(n\rightarrow  \gamma \chi)\approx \frac{g^2_s}{8 \pi} 4 \pi \alpha   \frac{ 8 \epsilon^2}{m_n^2} (m_n-m_{\chi})\approx 4 \times 10^{-27} (m_n-m_{\chi}),
\eeq
which almost the same with the $7\times 10^{-27}(m_n-m_{\chi})$  estimated in ref. \cite{FornalBej} to fit the free neutron life time. 
We thus see that the radiative width for a bound neutron is of the same order with that estimated  for a free neutron.
\section{Discussion and Remarks}
Our estimates of  the expected widths for baryon number violating neutron decay to a dark matter particle inside the nucleus are summarized in table \ref{tab:outdata} and Fig. \ref{fig:figbe11}. In Fig.  \ref{fig:figbe11} we have included the results obtained in the context of NCSMC, resulting from 
 an almost  ab initio constructed wave function, as well as those obtained in the context of the LST.
   It is encouraging that these two treatments  yield results, which essentially agree with each other, and both significantly enhance the  widths obtained in the context of the  HO shell model. \\
For a better understanding of Fig. \ref{fig:figbe11} we note that,  there appear two terms in the expression of the width, Eq. (\ref{Eq:widthbound}). The first is kinematical in nature and increases with the $\sqrt{m_n-B-m_{\chi}}$. The other comes from the form factor through the $\alpha$ dependence and as a result it decreases  with $m_n-B-m_{\chi}$. The two terms  tend in the opposite direction. In the region of interest to us the form factors in the case of NCSMC and LST models decrease very fast as a function of $\alpha$,  see Fig. \ref{Fig:facndecay}, and, as a result, for a given $B$ the width tends to decrease with  $m_n-m_{\chi}$. In the case of the HO model the form factor decreases much slower with $\alpha$ and the net result is a width, which slightly increases with $m_n-m_{\chi}$. 
 Note that the HO wave function does not extend far enough and, as a result,  the form factor gets suppressed, leading to  a smaller decay width.

 It thus  appears that the expected widths for baryon number violating neutron decay to a dark matter particle inside the nucleus are much larger than that expected in the case of the free neutron \cite{FornalBej}, provided that its mass is around 938 MeV.

  Note that the uncertainty due to the experimental error (1$\%$) is far below those due to the theoretical evaluations.
	
It is finally remarked that the experimental width for $^{11}$Be is smaller by factors around $10-3$ than that evaluated  for $n \rightarrow \chi$ with $m_{\chi} \le $ 939 MeV  in $^{11}$Be, and thus  the free neutron decay of $n\rightarrow \chi \gamma $ with $m_{\chi}$ = 937.90 Mev - 938.78 of ref. \cite{FornalBej}  can  hardly be the major channel to account for the neutron lifetime discrepancy between the "`bottle"' and "`beam"' experiments.

\section{Appendix evaluation of the nuclear matrix element}
We need the structure of the initial $A(N,Z)$ nucleus and the structure of the $A(N-1,Z)^*$ final nucleus. The essential information is contained in the CFP (coefficient of fractional parentage) 
$$\langle A(N-1,Z)J_f;j(n);A(N,Z)\rangle$$
which separates out the interacting neutron indicated by the quantum numbers $n_r\ell j$ and essentially gives the overlap involving the non interacting nucleons or spectroscopic factor. This can be obtained by a nuclear structure calculation or in some cases extracted from experiment in reaction involving a knock out neutron.
Then the matrix element involved is:
\beq
\mbox{ME}=\kappa_{\mbox{{\tiny scale}}}\, \langle A(N-1,Z)J_f;j(n);A(N,Z)\rangle \langle J_f M_i-m j m|J_i M_i\rangle J{me}
\eeq
where $J_{me}$ will be defined in Eq. (\ref{Eq:Jme}) below.

\subsection{The elementary transition  matrix element}
Let us suppose that the decaying nucleon is in a shell model state $n_r \ell j m ({\bf r})$, where $r$ is identified with the enter of mass coordinate of three quark system, i.e. 
 $${\bf r}\equiv{\bf R}_{cm}=\frac{1}{3}(\bf r_1+r_2+r_3)=\frac{{\bf R}}{\sqrt{3}}$$
and the outgoing dark mater particle is an spin state $m_s$. We must first evaluate the me
\beq
J_{me}=\langle n_r \ell j m|e^{i {\bf }q.{\bf r}}|m_s\rangle=4 \pi\sum_{\ell' m'}(i)^{\ell'}\langle n_r \ell j m|j_{\ell'}\left ({q r} \right ) Y^{\ell'}_{m'}(\hat{r})m_s\rangle \left (Y^{\ell'}_{m'}(\hat{q})\right )^*
\label{Eq:Jme}
\eeq
where $j_{\ell'}(z)$ is a spherical Bessel function and $ Y^{\ell'}_{m'}(\hat{r})$ the usual spherical harmonic. Using the standard angular momentum re-coupling we obtain a contribution only when $\ell'=\ell,\,m'=m_s-m$. i.e. :
\beq
 J_{me}=4 \pi f_{j,\ell} 2 \sqrt{2} a_N \sqrt{a_N}F_{n_r\ell}\left (\sqrt{2}q a_N\right )\langle j,m,\,\ell,m_s-m|1/2m_s\rangle  (-1)^{\ell}\left (Y^{\ell}_{m_s-m}(\hat{q})\right )^*,
\eeq
where $f_{j,\ell}$ is the needed angular momentum re-coupling factor given by
 \beq f_{j,\ell}=\left \{ \begin{array}{cc}1/\sqrt{2\ell+2},&j=\ell+1/2\\1/\sqrt{2\ell},&j=\ell-1/2\\ \end{array}\right .
\label{Eq:fjl}
\eeq
and 
$a_N$ is the nuclear harmonic oscillator size parameter and 
\beq
F_{n_r\ell}\left (\sqrt{{2}}q a_N\right )=\int_0^{\infty} x^2 dx \psi_{n_r\ell}(x)j_{\ell'}\left ({q \sqrt{2} a_N x }\right ), \,x=\frac{r}{a_N \sqrt{2}}
\label{Eq:NucFormFac}
\eeq 
(dimensionless ``form factor'').
\beq
F_{0s}(\alpha)=\frac{1}{2} \sqrt[4]{\pi }
   e^{-\frac{\alpha^2}{4}}
\eeq
\beq
F_{0p}(\alpha)=\frac{\pi^{1/4}}{2 \sqrt{6}}\alpha e^{-\frac{\alpha^2}{4}}   
\eeq
\beq
F_{0d}(\alpha)=\frac{\sqrt[4]{\pi }
   \alpha ^2 e^{-\frac{\alpha ^2}{4}}
   }{2
   \sqrt{15}}
\eeq
\beq
F_{1s}(\alpha)=-\frac{\sqrt[4]{\pi }
   e^{-\frac{\alpha ^2}{4}}
   \left(\alpha ^2-3\right)}{2
   \sqrt{6}}
\eeq
\subsection{The invariant amplitude squared}
The next step is to obtain $|\mbox{ME}|^2$ average over the initial m-sub-sates and sum over the final m-sub-states.
The result is 
\barr
\langle |\mbox{ME}|^2\rangle&=&\left (\kappa_{\mbox{{\tiny scale}}}\, \langle A(N-1,Z)J_f;j(n);A(N,Z)\rangle 4 \pi f_{j,\ell} 2 \sqrt{2} a_N \sqrt{a_N}F_{n_r\ell}\left (\sqrt{{2}}q a_N\right ) \right )^2\nonumber \\
 &&\frac{1}{2 J_i+1}\sum_{M_i,m}\langle J_f M_i-m j m|J_i M_i\rangle^2\langle j,m,\,\ell,m_s-m|1/2m_s\rangle^2Y^{\ell}_{m_s-m}(\hat{q})\left( Y^{\ell}_{m_s-m}(\hat{q})\right )^*\nonumber\\
&= &\left (\kappa_{\mbox{{\tiny scale}}}\, \langle A(N-1,Z)J_f;j(n);A(N,Z)\rangle 4 \pi f_{j,\ell} 2 \sqrt{2} a_N \sqrt{a_N}F_{n_r\ell}\left (\sqrt{{2}}q a_N\right ) \right )^2  \frac{2}{4 \pi}
\label{Eq:InvAmpSq}
\earr
\subsection{Neutron decay in the nucleus with photon emission}
We will now consider that the neutron before its decay emits a photon via its magnetic moment in a two step process, i.e.
$$ J_i\lim_{ n \ra \gamma n}J_n\,\lim_{n \ra \chi}J_f $$
In this case we have:
\beq
\Gamma \left (J_i\rightarrow J_f \chi \gamma\right)=\Gamma_{\gamma}\frac{1}{\delta E} \Gamma \left (J_i\rightarrow J_f \chi \right)
\eeq
where $\Gamma_{\gamma}$ the radiative width, $\Gamma \left (J_i\rightarrow J_f \chi \right)$ the width for neutron decay inside the nucleus discussed above and $\delta E$ an energy denominator,  essentially the photon energy.

The neutron photon interaction is given by
\beq
 H=\frac{g_s}{2 m_n} \sqrt{4\pi \alpha} ({\bf k}\otimes \sig)
\eeq
Thus one finds the corresponding invariant amplitude
\beq
{\cal M}(k)^2=\left (\frac{g_s}{2}\right )^2 4 \pi \alpha \frac{2}{3} \frac{k^2}{m^2_n} \langle||\sigma||\rangle^2 
\eeq
where $k$ is the photon momentum and  $\langle||\sigma||\rangle$ is the reduced $\mbox{ME}$ of the spin normalized to $\sqrt{6}$ for $s_{1/2}$-states.
We thus find for the  width for photon emission:
 \beq
  \Gamma_{\gamma}=\frac{1}{(2 \pi)^2} 4 \pi \int k^2 dk \frac{1}{2k}{\cal M}(k)^2 \delta{(\delta E-k)}=\frac{1}{\pi}\left (\frac{g_s}{2}\right )^2 4 \pi \alpha \frac{1}{3}\frac{ {(\delta E)^3}}{m_n^2} \langle||\sigma||\rangle^2 
	\eeq 
	(the factor $ 1/(2k)$ in the first expression comes from the photon normalization).

The total photon width for the neutron decay for photon emission is given by:
\barr
\Gamma \left (J_i\rightarrow J_f \chi \gamma\right)&=&\Gamma \left (J_i\rightarrow J_f \chi \right) \frac{1}{\pi}\left (\frac{g_s}{2}\right )^2 4 \pi \alpha \frac{1}{3}\frac{ {(\delta E)^2}}{m_n^2} \langle||\sigma||\rangle^2. 
\earr
we thus obtain for $1s_{1/2}$ neutron and $\delta E=m_n-m_{\chi}-B$:
\beq
\frac{\Gamma \left (J_i\rightarrow J_f \chi \gamma\right)}{\Gamma \left (J_i\rightarrow J_f \chi \right)}=\left (\frac{g_s}{2}\right )^2 \frac{2}{\pi} 4 \pi \alpha \frac{(m_n-B-m_{\chi})^2}{m_n^2}.
\eeq
Or
\beq
\frac{\Gamma \left (J_i\rightarrow J_f \chi \gamma\right)}{\Gamma \left (J_i\rightarrow J_f \chi \right)}\approx 2.4 \times 10^{-7}\frac{(m_n-B-m_{\chi})^2}{1 \mbox{MeV}^2}.
\eeq
\section{Single nucleon wave functions}
\label{sec:Localcale}
In this case one performs an isometric transformation \cite{karaAmos05} of the nucleon wave function $u(r)$ :
\beq
u(r)\rightarrow v(r)=s(r)u(f(r)),\, s(r)=\frac{f(r)}{r}\sqrt{\frac{df}{dr}}
\eeq
where $f(r)$ is a continuous function with the properties:
\beq
f(r)\rightarrow r \mbox{ as } r\rightarrow 0,\,f(r)\rightarrow \gamma \sqrt{r} \mbox{ as } r \rightarrow \infty,\,\gamma=2 a_N\frac{\sqrt{2m_n\epsilon}}{\hbar}
\eeq
where the parameter $\epsilon$ they employed  is the binding energy of the neutron considered as positive. i.e. $B$ in our notation. Thus this transformation achieves
\beq
e^{-\frac{r^2}{2  a_N^2}}\rightarrow e^{-r \left(\frac{\sqrt{2m_n\epsilon}}{\hbar}\right )} \mbox{ for large } r
\eeq
Thus it changes a Gaussian into a Yukawa like behavior. In the case of $^{11}$Be  a binding energy of 0.8 MeV was previously employed  \cite{karaAmos05}. In the present calculation we will employ the more up to date value  of $B=$0.51 MeV.
The authors found it appropriate to employ the following function:
\beq
f(r,m,\gamma)=\left[\left(\frac{1}{\gamma 
   \sqrt{r}}\right)^m+\left(\frac{1}{r}\right)^m\right]^{\left(-\frac{1}{m}\right )}
\eeq
with $m=4$ and $m=8$. It turns out that for our purposes it does not make much difference which of the two values is adopted.
\end{document}